\documentclass[12pt]{article}
\input epsf
\usepackage[utf8]{inputenc}
\usepackage{graphicx} 
  \expandafter\let\csname equation*\endcsname\relax
  \expandafter\let\csname endequation*\endcsname\relax
\usepackage{amsmath} 
\usepackage{amssymb} 
\usepackage{euscript}
\usepackage{mathrsfs}

\begin{document}

\title{Seeking for the observational manifestation of de Sitter Relativity}

\author{D.A. Tretyakova\footnote{E-mail: daria.tretiakova@urfu.ru.}, \\
Department of Theoretical Physics, Physics Department, \\Institute for Natural Sciences, Ural Federal University, \\Lenin av. 51, Ekaterinburg, 620083, Russia}

\date{\today}

\maketitle

\begin{abstract}
The de Sitter invariant special relativity is a  natural extension of the usual Einstein special relativity. Within this framework a generalization of special relativity (SR) for the de Sitter space-time introduces a new length scale $R$, serving an origin of geometrical cosmological constant $\Lambda=3/R^2$. De Sitter relativity predicts the departure from the Lorentz invariance due to space-time curvature, related to the geometrical cosmological constant. In this paper the possible impact of de Sitter special relativity effects on threshold particle processes and equivalence principle violation is considered. The main conclusion is that constraints, coming from cosmological fine structure constant variations render this effects nowadays undetectable. A brief outlook is given thereafter.
\end{abstract}

\section{Introduction}
\label{s0}

A large variety of astrophysical data sets ranging from high redshift surveys of supernovae to WMAP observations  indicate that our Universe experiences an accelerating phase of expansion \cite{2013ApJS19H}. A possible interpretation of this expansion in terms of General Relativity (GR) states that about $70\%$ of the total energy of our Universe is attributed to the dark energy  with large and negative pressure, for which the cosmological constant $\Lambda$ is considered as the best fit nowadays (see \cite{Bamba:2012cp} for a comprehensive review on dark energy ore \cite{2013FrPhyL} for a brief one). 
Another reason to consider $\Lambda$ in the GR action is that it is a rightful term along with the scalar curvature, so neglecting it is a sort of fine-tuning.
However the origin of the cosmological constant remains unknown.

A way to substantiate the cosmological constant and account for it is given in de Sitter Relativity (dS-SR) \cite{Aldrovandi:2006vr}. Within this framework a generalization of Special Relativity (SR) for de Sitter space-time introduces a new length scale $R$, serving an origin of geometrical cosmological constant $\Lambda=3/R^2$.
 The Poincare group, being the full symmetry group of any relativistic field theory, is then replaced with de Sitter group. The de Sitter special relativity can be viewed as made up of two different relativities: the usual one, related to translations, and a conformal one, related to proper conformal transformations, interpolating between these two limiting cases. Usual SR energy-momentum-relation and transformation rules restore for $R\to\infty$, so for small values of $\Lambda$ the Poincare symmetry will be weakly deformed \cite{Aldrovandi:2006vr}. 

DS-SR is usually said to be a specific kind of deformed (or doubly) special relativity \cite{AmelinoCamelia:2000ge}. The difference from doubly special relativity however is significant, for example the  time-dependence of the dispersion relation and the absence of a ``rainbow'' - the wavelength dependence of the speed of light. 

Another important application of de Sitter group emerges in high energy physics. The idea that the curvature of the Universe must tincture  physical processes at energies high enough seems reasonable.   Once  we consider the phase transitions from the spontaneously broken symmetries as the primary source of a non-vanishing $\Lambda$ (vacuum energy), it is also reasonable to assume that a high energy event could modify the local structure of space-time for a short period \cite{Mansouri:2002cg}. In this case the Minkowski space-time would turn into a de Sitter one in the vicinity of a high energy collision. This local de Sitter geometry would have nothing to do with dark energy, it would be sourced by the energy-momentum of the system. According to this point of view, there would be a connection between the energy scale of the experiment, the local value of $\Lambda$ and the corresponding Lorentz-invariance violation measure. 
For an experiment with energy of the order of the Planck one, the local value of the cosmological constant would be  $\Lambda \sim 10^{66} cm^{-2}$ , which differs from the  observed  cosmological constant by roughly 120 orders of  magnitude.  A very peculiar new quantum world would then emerge, whose physics has yet to be developed.
All the above makes the consideration of de Sitter relativity prudent.

The purpose of this paper is to consider the relevance and detectability of dS-SR effects for terrestrial experiments. Section \ref{s1} gives a brief introduction to dS-SR kinematics. In section \ref{s1} we also briefly reconsider the existing bounds on dS-SR in the scope of recent experimental and observational data and determine the strongest constraint.  In section \ref{s2} we explore the impact of dS-SR on threshold reactions, in particular, GZK photopion production. Section \ref{s3} is devoted to the equivalence principle violation. A brief conclusion is given thereafter.

\section{De Sitter Relativity} \label{s1}
De Sitter relativity starts with an embedding of a 4-hypersurface $S_R$ in a 5d pseudoeuclidean space
\begin{eqnarray}\label{271}
 S_R:\eta_{AB}\chi^A \chi^B=-R^2.
 \end{eqnarray}
Via the gnomonic projection,  displaying all great circles as straight lines (resulting in any straight line segment on a target spacetime showing a geodesic) $S_R$ gives rise to the Beltrami metric for 4-coordinates $x^\mu$:
\begin{eqnarray}\label{star281}
 B_{\mu\nu}(x)&=& \frac{\eta_{\mu\nu}}{\sigma (x)}
+\frac{\eta_{\mu\lambda}\eta_{\nu\rho} x^\lambda x^\rho}{R^2 \sigma(x)^2}, \\
\sigma(x)  &\equiv & 1-\frac{1}{R^2}
\eta_{\mu\nu}x^\mu x^\nu, ~~
\eta_{\mu\nu} =diag(1,-1,-1,-1).
\end{eqnarray}
which acts as the inertial frame system in dS-SR \cite{Sun:2013vfa}.

According to ref. \cite{Yan:2005wf}, the Lagrangian for a free particle in dS-SR reads
\begin{eqnarray}\label{271}
 L_{{dS}}(t,x,\dot{x})=-m_0c\frac{ds}{dt}
 =-m_0c \frac{\sqrt{B_{\mu\nu}(x)dx^\mu dx^\nu}}{dt}&& \\
 =-m_0c \sqrt{B_{\mu\nu}(x)\dot{x}^\mu \dot{x}^\nu},&&
 \end{eqnarray}
where $\dot{x}^\mu=\frac{d}{dt}x^\mu$.
Here the speed of light parameter $c$ and the radius $R$ of the pseudo-sphere in de Sitter space are two universal constants in the theory. 
The following relation between $R$ and the geometrical cosmological constant exists:
\begin{eqnarray}\label{RL}
&&\Lambda=3/R^2.
\end{eqnarray}

The existence of inertial coordinate system, the one in which the inertial motion law for free particles holds, is the keystone of special relativity. So the fact, that the Beltrami metric is inertial  allows one to build up the new special relativity: dS-SR. 
One immediate consequence of dS-SR is that the familiar dispersion relation $E^2=m^2c^4+p^2c^2$ is modified to give the following expression:
\begin{eqnarray}\label{dp}
E_{dS}^2 &=& m_0^2 c^4+{ p}_{dS}^2 c^2 + \frac{c^2}{R^2}
({ L}_{dS}^2-{ K}_{dS}^2), 
\\
p_{dS} &=& m_0 \Gamma \dot{x},
\\
 E_{dS} &=&  m_0 c^2 \Gamma,
\\
\label{phys_mom} K_{dS} & =& m_0 c \Gamma (x -t\dot{x})=m_0c\Gamma x-ctp_{dS},
\\
 L_{dS} & =& -m_0 \Gamma \epsilon_{\;jk} x^j
\dot{x}^k=-\epsilon_{\;jk} x^jp^k_{dS}.
\end{eqnarray}
Here the Noether charges $E_{dS}, p_{dS},L_{dS},K_{dS}$ are physical energy, momentum, angular-momentum
and boost charges respectively. The analog of the Lorentz factor $\gamma$ is:
 {\begin{eqnarray} \nonumber
 && \Gamma^{-1}\hskip-0.1in =\sigma(x) \frac{ds}{c dt} \\
&& \!\!\!\!\!\!\!\!\!\!\!\!\!\! =\frac{1}{R} \sqrt{(R^2-\eta_{ij}x^i
x^j)(1+\frac{\eta_{ij}\dot{x}^i \dot{x}^j}{c^2})+2t \eta_{ij}x^i
\dot{x}^j -\eta_{ij}\dot{x}^i \dot{x}^j t^2+\frac{(\eta_{ij}
x^i\dot{x}^j)^2}{c^2}}. \label{new parameter}
\end{eqnarray}}
When $|R| \to \infty$, $\Gamma \to \gamma$.
One immediately notes that \eqref{dp} depends on cosmological time $t$, Noether charges of Lorentz boost and rotations in space
(angular momenta), and hence it  depends on a choice of the space-time origin. A natural choice of such an origin would be the Big Bang occurrence. Then  terrestrial experiments occur at $t_0 \approx 13.7Gy.$ and $x_0\equiv x(t_0)\approx 0$ \footnote{For example, the distance between CERN and OPERA  is nearly $731$ km, corresponding to the path and flight time $ x \sim 10^{-23} R$, $ t \sim 10^{-21} t_0$, hence $ t \approx 0$, $ x / R \approx 0$ }. This gives the following expressions for the Noether charges and the dispersion relation
\begin{eqnarray}
 K&\simeq& -ct_0p_{dS} , \quad L_{dS}\simeq 0, \label{disp1} \\
E_{dS}^2 &= &m_0^2 c^4+{ p}_{dS}^2 c^2\left (  1- \frac{c^2t_0^2}{R^2} \right ). \label{disp2}
\end{eqnarray}
Let us introduce for further simplicity
\begin{eqnarray}
&& \lambda^2=c^2t_0^2/R^2.
\end{eqnarray}

Thus dS-SR reports a very special kind of Lorenz-invariance violation, evolving with cosmic time.  The effects, predicted by dS-SR are the following.
\emph{Length scale.}
The de Sitter transformations can be thought of as rotations in a five-dimensional pseudo-Euclidian space-time. Since these transformations leave invariant the quadratic form $S_R$ they also leave invariant the length parameter $R$ \cite{Aldrovandi:2006vr}.
The very existence of a relativistic-invariant fundamental length may have interesting consequences and provide an alternative procedure of field theory regularization. 


\emph{Superluminal motion of massive particles}. The speed of light in dS-SR remains unaffected, particles however can potentially move faster than light \cite{Yan:2011jm}. This opportunity provided by dS-SR was used to explain OPERA superluminal neutrinos \cite{Yan:2011ea}. Taking the OPERA neutrino flight trajectory as $\{x^1\equiv x(t), \;x^2=0,\; x^3=0\}$ we have:
\begin{eqnarray} \label{Energy1}
v_{dS}\equiv \dot{x}(t)&=&\frac{c^2 p_{dS}}{E_{dS}}, \\
\label{energy2} E_{dS} &=&\frac{m_0 c^2}{\sqrt{1-(\frac{v_{dS}}{c})^2+(\frac{x_0-v_{dS}t_0}{R})^2}},
\end{eqnarray}
and then obtain the neutrino velocity
\begin{eqnarray}
&& v_{dS}=c \sqrt{1-m_0^2 c^4/E_{dS}^2 \over 1-\lambda^2}.
\end{eqnarray}
 However, since the effect was disclaimed, the estimate of the cosmological constant given in \cite{Yan:2011ea} may be turned into a constraint, stating that no superluminal motion is observed by now. According to \cite{Adam:2012pk} the neutrino speed bound is $-1.8 \times 10^{-6} < (v_{\nu}-c)/c < 2.3 \times 10^{-6}$ at $90\%$ C.L., hence
\begin{eqnarray} \label{r_est}
&&R>6.39 \times 10^{3} Gl.y.
\end{eqnarray}
The absence of a bremsstrahlung for superluminal neutrinos in dS-SR was shown in \cite{Yan:2011jm,Li:2011ad,Hu:2012gd}.

\emph{The evidence of non-geometrical cosmological constant}. Indeed if we assume that $R$ is the only source for the cosmological constant, then, using \eqref{RL} and modern WMAP data \cite{2013ApJS19H} we obtain:
\begin{eqnarray}\label{L1}
&&\Lambda=\cfrac{3 H_0^2}{c^2} \Omega_{\Lambda 0}, \qquad R\approx 17 Gl.y., 
 \end{eqnarray}
which is incompatible with the 
constraint above. Hence the maximum geometrical input into $\Lambda$ can be estimated as
\begin{eqnarray}\label{L2}
&&\Lambda_{R}\leq 10^{-56} m^{-2}, 
 \end{eqnarray}
which is \emph{O(e-4)} compared to the observed value. This indicates that within the framework of dS-SR the observed cosmological constant should have a dynamical origin. 


\emph{Fine structure constant spatial and temporal variation}, which was reported for dS-SR in \cite{2011ChPhC..35..228Y}. Modern theories directed toward unifying gravitation with the three other fundamental interactions suggest variation of the fundamental constants in an expanding universe, so the general idea is widely considered. 
The fine atomic spectrum structure of distant galaxies, compared to the laboratory spectrum can tell us whether fine structure constant varies with time.
An observational signature of the fine-structure constant $\alpha$ spatial and temporal variations was discovered in \cite{Murphy:2000nr, Webb:2010hc}. The temporal variation, reported in \cite{Murphy:2000nr} corresponds to approximately $\dot{\alpha}/\alpha\approx 7\times 10^{-17} yr^{-1}$ while modern bounds from terrestrial experiments reported in \cite{PhysRevLett.113.210801} state 
\begin{eqnarray}\label{al_bound}
&&\dot{\alpha}/\alpha\leq-0.7(2.1)\times 10^{-17} yr^{-1}, 
\end{eqnarray}
more in agreement with another cosmological data from \cite{PhysRevLett.92.121302}. 
According to \cite{2011ChPhC..35..228Y} for a hydrogen atom of a distant source one obtains
\begin{eqnarray} 
&&\cfrac{\Delta\alpha}{\alpha}=-\cfrac{c^2t^2}{2R^2},
\end{eqnarray}
where $t$ is the difference in the cosmological time between the Earth and the source. Given the bound \eqref{al_bound} we can estimate
\begin{eqnarray} \label{r_est1}
&& R\geq 1.6 \times 10^{4} \mbox{Gl.y.}
\end{eqnarray}
The spatial variation of $\alpha$ can also be reproduced within dS-SR \cite{Feng:2015vla}. Testing this spatial variation  indicated by Webb et al.  with  laboratory atomic measurements requires at least $\dot{\alpha}/\alpha=10^{-19}yr^{-1}$ sensitivity which may be achieved  soon \cite{Safronova:2014ela}.


Thus we briefly reconsidered the existing bounds on dS-SR in the scope of recent experimental and observational data, basically improving them by an order of magnitude. In what follows we turn to the scope of this paper: consider the impact of dS-SR formalism on the GZK threshold reaction.

\section{Threshold reactions} \label{s3}

Threshold particle collisions represent a well-known testbed of relativity. Collisions at the threshold are also very special, since they force all inertial observers to agree whether the  kinematic balance of  the threshold reaction is satisfied. Hence, kinematic conditions must hold under the appropriate relativistic transformations. Ref. \cite{AmelinoCamelia:2000mn} shows that for the deformed dispersion relations the kinematic conditions may deform as well, thus shifting the threshold. So it is necessary to check whether it happens in dS-SR before applying such conditions to specific processes. For the dispersion relation 
\begin{eqnarray}  \label{dr2}
 &&E_{dS}^2 =m_0^2 c^4+{ p}_{dS}^2 c^2\left (  1- \lambda^2 \right ), 
\end{eqnarray}
following \cite{AmelinoCamelia:2000mn} we make the ansatz for the boost along the particle propagation direction
\begin{eqnarray} 
 && B=i[cp+\Delta_1(E,p)]\cfrac{\partial}{\partial E}+i[E/c+\Delta_2(E,p)]\cfrac{\partial}{\partial p}, 
\end{eqnarray}
describing infinitesimal transformations. For the deformation functions \\ $\Delta_{1,2}\to 0$ we  arrive at the  familiar differential boosts generators of Lorentz group. To preserve the dispersion relation we fix the boost generator as
\begin{eqnarray} 
 && B=icp\cfrac{\partial}{\partial E}+i\left [\cfrac{E}{c(1-\lambda^2)}\right ]\cfrac{\partial}{\partial p}. 
\end{eqnarray}
 This form of the boost implies consistency of two inertial observers descriptions of the same particle motion. The corresponding infinitesimal transformations are
\begin{eqnarray} \label{inf_tr}
 && \cfrac{\partial E}{\partial \xi}=-cp, \qquad
\cfrac{\partial p}{\partial \xi}=-\cfrac{E}{c(1-\lambda^2)}, 
\end{eqnarray}
with $\xi$ being the familiar boost (rapidity) parameter. Let us further discuss  the simple case of a scattering process $\mathit{a+b \to c+d}$ (collision processes with incoming particles {\it a} and {\it b} and outgoing particles {\it c} and {\it d})\footnote{We adopt here one-dimensional consideration describing a head-on {\it a - b} collisions producing {\it c - d} at threshold, when the particles produced have no energy available for the momentum components in the  orthogonal directions.}. One can easily see that the kinematic conditions for this threshold process
\begin{eqnarray} 
 && E_a+E_b-E_c-E_d=0, \\
&& p_a+p_b-p_c-p_d=0
\end{eqnarray}
hold under \eqref{inf_tr}, so the threshold conditions in dS-SR in the vicinity of Earth remain the same as for SR.

An example of the observed threshold reaction is a photopion production, forming the cosmic rays GZK cut-off \cite{1966JETPL...4...78Z, PhysRevLett.16.748}. Ultra high energy (UHE) protons  loose energy due to  the interaction with CMB photons and should slow down until their energy is below the GZK energy $5\times 10^{19}eV$. 
\begin{eqnarray}
&& p^++\gamma_{CMB}\to \Delta^+ \to p^++\pi^0, \\
&& p^++\gamma_{CMB}\to \Delta^+ \to n+\pi^+.
\end{eqnarray}
A sufficiently energetic CMB photon, emerging at the tail of the spectrum,
is seen in the rest frame of an UHE proton having the energy above  the threshold $140 eV$ for pion production, so the proton’s mean free path through the CMB decreases
exponentially with energy (down to a few Mpc) above the GZK limit. Hence, since there are no sources of UHE cosmic rays close enough to Earth, UHE protons should not be detected. 
The observational situation regarding the GZK suppression is controversial. There are data sets both confirming the GZK cut-off \cite{Abbasi:2007sv,Abraham:2008ru} and conflicting with it \cite{PhysRevLett.81.1163}. So the issue is considered open and the mechanisms, extending the  UHE cosmic ray horizon, are widely considered \cite{Olinto:2000sa,AmelinoCamelia:2000zs}. One of the possibilities is the threshold momentum to shift itself due to a change in the dispersion relation \cite{AmelinoCamelia:2000zs}, what we are about to find for dS-SR. Considering the GZK limit in the context of dS-SR,  we get from \eqref{dr2}:
\begin{eqnarray}  \label{dr3}
 E_{i} &\approx&  p_{i} c\left (  1- \cfrac{\lambda^2}{2} \right ) +\cfrac{m_0 c^2}{{ p}_{i} c (  1- \lambda^2/2)}. 
\end{eqnarray}
For the GZK process we assume particle {\it a} being a proton, particle {\it b} being a CMB photon. 
We obtain for the proton momentum in the laboratory frame
\begin{eqnarray}
p_{dS} &\approx &\cfrac{(m_c+m_d)^2-m_{p^+}^2}{4 \epsilon_{dS}}\,c^3, \\
\epsilon_{dS} &\approx & p_{\gamma_{CMB}}c\left(  1- \cfrac{\lambda^2}{2} \right ),
\end{eqnarray}
with $\epsilon_{dS}$ being the CMB photon energy. In general $\epsilon_{dS}<\epsilon_{SR}$ hence $p_{dS}>p_{SR}$, so the de Sitter curvature extends the high energy cosmic ray cut-off. However the threshold momentum has to increase up to a factor of 6 at least to explain the GZK violation, reported by AGASA project, which is impossible, since $\lambda^2 \approx 10^{-24}$ according to \eqref{r_est1}. So the data, disclaiming the GZK paradox agree with dS-SR.

Ref.  \cite{Yan:2005wf} shows that the dS-SR the free particle Hamiltonian is time- and coordinate-dependent, so one might wonder whether this would impact a particle moving through distances above $160 Ml.y.$ However the Noether theorem assures the corresponding charges to be conserved, thus, the energy conservation law and the dispersion relation  in dS-SR hold, despite the fact that the dS-SR dynamics is ruled by a time-dependent Hamiltonian. Hence the threshold condition must hold through the path of the particle.


\section{Equivalence principle} \label{s2}
Finally we would like to give some notes on using the equivalence principle violation to constrain dS-SR. It is known  that Lorentz invariance violation implies a violation of the equivalence principle (EP) \cite{lrr-2005-5}. We can examine this statement for dS-SR and estimate the corresponding EP violation. In the neighborhood of earth conditions \eqref{disp1} and \eqref{disp2} obviously hold since $ x \sim 10^{-20} R$ for the moon orbit. If we assume Hamiltonian dynamics at low energy and use the energy as the Hamiltonian, then for a non-relativistic particle in a weak gravitational field $\Phi(x)$ we have 
\begin{eqnarray}
&& H \approx m_0c^2 +m_0\Phi(x)+\cfrac{ p^2}{2m_0}\left (  1+ \lambda^2 \right )+ \cfrac{c^2}{2R^2}(m_0x_0^2-2x_0t_0p)\\
&& \ddot{x}=-\cfrac{\partial \Phi}{\partial x} \left (  1+ \lambda^2 \right ). \label{eqp}
\end{eqnarray}
The  factor on the rhs of  \eqref{eqp}does not distinguish different bodies, so the EP is not violated (up to the precision at which $x=x_0$). However EP breaks down when time and coordinate dependence of the dispersion relation becomes valid. At this point the acceleration \eqref{eqp} may in general depend on mass, time, coordinates and momentum. This can potentially be applied to wide enough  double pulsar systems where the weak field approximation for the gravitational field works and the PPN formalism applies. Special relativistic effects, such as time dilation and length contraction are observed in pulsars, hence, the dS-SR formalism is relevant. However for the dispersion relation \eqref{dp} we obtain
\begin{eqnarray}
&& \ddot{x}=-\cfrac{\partial \Phi}{\partial x} + \delta, \label{eqp2}\\
&&\delta=\cfrac{1}{2m_0R^2} \cfrac{d }{dt} \left [ \cfrac{\partial }{\partial p} \left({ L}_{dS}^2-{ K}_{dS}^2\right)\right].
\end{eqnarray}
Pulsar experiments constrain the difference $\delta_{psr}-\delta_{companion}$ \cite{Shao:2016ubu}. This difference, being small by itself from a point of view of the terrestrial observer in dS-SR (since the double  system components are far from us, close to each other and approximately equal in mass) is further suppressed by the factor $R^{-2}$. So the perspective of using double pulsars to constrain $R$ seems faint.

\section*{Conclusions}
The de Sitter Relativity is an interesting framework with both theoretical and practical implications. In this paper we considered the impact of dS-SR kinematics on the threshold reactions, in particular, GZK cut-off.  The results of section \ref{s1} together with results from the sections \ref{s2},\ref{s3} suggest that terrestrial experiments are nowadays unable to detect dS-SR effects.   
The most severe restriction on de Sitter relativity comes from the cosmological variation of the fine structure constant stating $R\geq 1.6 \times 10^{4} \mbox{Gl.y.}$. This constraint ensures that the de Sitter curvature does not affect the GZK threshold ore the equivalence principle essentially. 
The maximum geometrical input into the cosmological constant value can be estimated as
\begin{equation}\label{L2}
\Lambda_{R}\leq 10^{-58} m^{-2}, 
 \end{equation}
which is \emph{O(e-6)} compared to the observed value $\Lambda_{\Lambda CDM}\approx 10^{-52} m^{-2}$. This  indicates that within the framework of dS-SR the observed cosmological constant should have a dynamical origin. This conclusion is in conformity with with the result of \cite{2013ApJS19H}, stating that the equation of state parameter for the dark energy is $w_{DE_0} = -1.17(+0.13 -0.12)$ for the flat Universe at the $68\%$ C.L., disagreeing with purely cosmological constant dark energy $w_{DE_\Lambda}=-1$. 

One might suggest to use cosmic ray shower muon experiments,  measuring  energy and momentum separately thus  exploring the dispersion relation, as a testbed for dS-SR. However any value of a selected SR effect near the earth can be fitted by properly choosing $R$ for a given $t_0$, and due to the smallness of the correction factor $\lambda^2 \approx 10^{-24}$, any terrestrial experiment by itself seems nearly useless for constraining dS-SR (neutrino observations constrain $\lambda^2\leq 10^{-12}$ \cite{Bi:2011nd}).  Cosmological restrictions appear stiffer. The promising way to seek for an evidence of dS-SR might then be to consider the drift of the SR effects with cosmic time (i.e. with cosmic distance or red shift $z$) using $Ly-\alpha$ forest ore distant quasars. Viable tests must then be constructed as a combination of cosmological and high energy physics ones.

Another interesting tool to test the dS-SR formalism appears to be the equivalence principle violation. However the perspective of using double pulsars to constrain $R$ seems faint due to the specification of observational output. Double systems, containing a black hole and a pulsar could provide a better EP violation test once discovered \cite{Bagchi:2014tma},  a detailed calculation for this situation will be given elsewhere.


\bibliographystyle{unsrt}
\bibliography{mybib_eng}

\end{document}